\documentclass[preprint,showkeys,preprintnumbers,amsmath,amssymb]{revtex4-1}

\usepackage{graphicx}
\usepackage{dcolumn}
\usepackage{bm}

\begin{document}

\title{\emph{Z}-scanning Laser Photoreflectance as a Tool for Characterization of Electronic Transport Properties}

\author{Will Chism}

\email{email address: wchism@xitronixcorp.com}
\affiliation{Xitronix Corporation\\ 106 E. Sixth St., Ninth Floor, Austin, TX, USA 78701}

\date{\today}

\begin{abstract}

The physical principles motivating the \emph{Z}-scanning laser photoreflectance technique are discussed.  The technique is shown to provide a powerful non-contact means to unambiguously characterize electronic transport properties in semiconductors.  The technique does not require modeling of charge transport in the sample or a detailed theoretical model for the sample physics.  Rather, the measurement protocol follows directly from the simple relation describing the radial diffusion of carriers injected by a laser source. The use of a probe laser beam permits an analytic parameterization for the \emph{Z} dependence of the photoreflectance signal which depends soley on the focal parameters and the carrier diffusion length. This allows electronic transport properties to be determined with high precision using a nonlinear least squares fit procedure.  The practical use of the technique is illustrated by the characterization of carrier transport properties in semiconducting \emph{p-n} junctions.

\end{abstract}

\maketitle

\section{\label{sec:level1}Introduction}

The measurement of electronic transport properties in semiconductors, namely carrier diffusion lengths, recombination lifetimes, and mobilities, is a long standing problem in many areas of physics and engineering.  Moreover, evaluation of these parameters is essential in the semiconductor manufacturing industry because semiconductor device performance directly depends upon these electronic transport properties \cite{Schroder}. The techniques used to determine carrier transport properties may be divided into two classes: those requiring direct contact with the sample such as four-point probe (4PP) and Hall effect measurements; and  non-contact methods such as surface photovoltage (SPV) \cite{SPV,SPV2}, photoconductance decay \cite{FCAetc}, photoluminescence \cite{PL}, and time-resolved THz spectroscopy \cite{THz,Heilweil}. The main advantage of the contact techniques is their standardization \cite{ASTM}. However, direct contact with the sample risks damage and/or contamination, and can lead to misleading results.  For example, conventional 4PP techniques damage the sample and suffer from inaccuracies caused by over-sampling of the underlying substrate.  These risks are avoided by non-contact techniques. However, the interpretation of data obtained in non-contact measurements often involves detailed theoretical assumptions about the sample.

In this paper, the recently introduced \emph{Z}-scanning laser photoreflectance technique \cite{arXiv} is discussed and the technique is applied to the measurement of electronic transport properties in semiconducting \emph{p-n} junctions. It is shown the technique provides a powerful non-contact means to evaluate electronic transport properties while keeping theoretical assumptions to an absolute minimum. Photoreflectance (PR) is a particularly convenient type of modulation spectroscopy, as it may be performed at room temperature and only requires the sample have a reflecting surface \cite{Aspnes80}. In PR, an intensity modulated pump light beam is used to photo-inject charge carriers in a semiconductor sample, thereby modulating one or more physical quantities (\emph{e.g.}~the internal electric field), resulting in modulation of the sample reflectivity.  A second CW probe light beam, coincident with the pump beam, is used to detect the modulated reflectivity of the sample. The pump light is modulated at a known frequency so that a phase-locked detection circuit may be used to suppress unwanted noise, resulting in the ability to detect reflectance changes at the ppm level.  The measurement signal consists of a vector characterized by an amplitude and a phase.  The amplitude is the induced AC change in reflectance, often normalized by the DC reflectance, whereas the phase is the relative lag of the reflectance change with respect to the driving phase due to the interaction dynamics of photo-injected carriers within the sample. PR spectroscopy has been used to determine semiconductor bandstructures, internal electric fields, carrier concentration, and other material properties such as crystallinity, alloy composition, and physical strain \cite{Pollak}. Laser photoreflectance (LPR) techniques comprising the use of a laser probe beam in the conventional PR apparatus are generally well-suited for use in semiconductor device manufacturing as they are non-contact and can be performed rapidly with micrometer scale resolution \cite{JVST}.

Under the experimental conditions used for conventional PR, the signal arises from an electro-modulation effect \cite{Bottka,Shen90}.  However, thermal and free carrier effects on the optical properties may also provide significant contributions to the PR signal \cite{Opsal87}. In general the PR signal may be expressed as a superposition of electromodulation, carrier-modulation, and thermomodulation components:
\begin{eqnarray}
\Delta R = \frac{\partial R}{\partial E}\Delta E + \frac{\partial R}{\partial N}\Delta N +
\frac{\partial R}{\partial T}\Delta T,
\label{eq:drgen}
\end{eqnarray}
where $\Delta R$, $\Delta E$, $\Delta N$, and $\Delta T$ are the local variations in reflectance, electric field, carrier density, and temperature, respectively.  The vectorial nature of the electric field increases the diagnostic value of electromodulation relative to carrier- and thermomodulation \cite{Seraphin}. However, since carriers in a typical laminar structure are free to flow in directions parallel to the surface (and therefore will continue to flow until any transverse fields become zero), the electric field in a free-standing sample is usually in the direction normal to the surface. Thus PR is typically a realization of surface-barrier electroreflectance \cite{Seraphin,Aspnes73a}.  The polarization dependencies of the relevant coefficients of reflectance ($\partial R/\partial E$, $\partial R/\partial N$, and $\partial R/\partial T$) have been discussed in the literature \cite{Aspnes76,Wagner}.

More importantly, the dispersions of these reflectance coefficients determine the modulation components that can be realized at a given probe beam wavelength.  For example, the electromodulation coefficient arises from a third-order nonlinear susceptibility and exhibits a sharp third-derivative lineshape centered on the critical points of the semiconductor bandstructure \cite{Aspnes80,Aspnes73a}.  Thus the electromodulation component of the PR signal can be enhanced by selecting the probe beam wavelength at or very near the direct interband transitions of semiconductor material in the sample. The dispersion of the thermomodulation component is also centered at the critical points of the semiconductor bandstructure, but exhibits a relatively broad first-derivative lineshape \cite{Aspnes80,Batz}.  Thus, as a practical matter, a range of wavelengths in the UV-VIS region of the spectrum are suitable for detection of thermomodulation.  Of course, at or near the critical points either the electromodulation or thermomodulation component may be suppressed by a zero-crossing of the respective coefficient \cite{Seraphin,Batz}. On the other hand, the free-carrier coefficient is generally small in the UV but is proportional to the square of the wavelength \cite{Opsal87,Wagner}. Accordingly, the carrier-modulation signal is enhanced at wavelengths in the NIR-IR.  In addition to determining the particular physical quantities that may contribute to the PR signal, the probe beam wavelength determines the depth over which such physical quantities are sampled.  In particular, the PR signal is weighted by $\sim\exp\{-z/\delta\}$, where $z$ is the distance into the sample and $\delta$ is the probe absorption depth \cite{Aspnes69,Aspnes73,Wagner}. Thus variations in local physical quantities occurring beyond the probe absorption depth cannot directly affect the PR signal.

Other considerations such as pump beam wavelength, intensity, and modulation frequency are also important \cite{Behn,Shen88,Shen91}. This is due to the fact that the variations in physical quantities appearing in the r.h.s.~of Eq.~(\ref{eq:drgen}) are generally not independent. For example, photo-injected carriers with energies above the band edge will quickly thermalize with the lattice via transitions to the bottom of the conduction band \cite{Opsal85}.  Thus the use of pump beam wavelengths close to the band gap will operate to minimize the induced temperature variations.  Prior to recombination, an excess electron-hole density will exist in the semiconductor \cite{Opsal85}.  If the modulation period is shorter than the recombination lifetime the semiconductor cannot relax to equilibrium.  Thus the use of high modulation frequencies will generate a steady state excess carrier density.  Such an excess carrier density will also reduce any internal electric field present in the semiconductor \cite{Bottka,Shen90}. The use of high pump (or probe) intensities may similarly offset the sample from equilibrium.  These considerations are known in the art and have been provided to place the present work in context.  Specifically, it should be recognized that provided the experimental parameters are properly chosen in view of the semiconductor system under test and the physical parameters of interest, the principles of the \emph{Z}-scanning LPR technique discussed here are wholly applicable to the determination of transport properties associated with any of the modulation components appearing in Eq.~(\ref{eq:drgen}).

Despite the long-standing recognition that, at least in principle, LPR techniques may be used to determine carrier and/or thermal transport properties \cite{SMR,Opsal85,Opsal83}, LPR is not widely used for that purpose.  At the outset, determination of transport properties from PR or other more commonly used measurement techniques such as SPV generally involves a theoretical model of the sample physics.  For example, analytic expressions for $\Delta V$ (or $\Delta N$) are typically obtained from the solution of a one-dimensional (1D) differential equation for electronic transport in the sample \cite{SPV2}. Provided the measurement signal has an analytic relation with $\Delta V$, transport properties may then be determined from the measurement. However, only in certain limiting cases are the desired transport properties directly related to the measured quantities (\emph{e.g.}~SPV amplitude and phase).  Furthermore, in the three-dimensional (3D) limit, which is encountered when tightly-focused Gaussian laser beams are used, the excess carrier density involves a Hankel transform of the 1D solution and therefore must be treated numerically \cite{Mandelis}. This exacerbates the problem of determining transport properties by requiring case-specific numerical analyses. A more direct approach involves measuring the LPR signal as a function of separation of pump and probe beams such that lateral diffusion may be spatially resolved \cite{SMR}.  However, the requirement of precisely controlling and metering the distance between laser spots complicates the experimental design and operation.  This has prevented the industrial use of such transverse scanning techniques to date.  An alternative suggestion involves holding the pump-probe offset fixed and recording the LPR phase with respect to the modulation frequency \cite{Idaho}, since the LPR signal will also depend on modulation frequency due the dependence of diffusion length on modulation frequency \cite{SMR}.  However, the determination of the optimal pump-probe offset for this type of measurement remains an empirical challenge.

\section{The \emph{Z}-scanning LPR Technique}

The \emph{Z}-scanning LPR technique provides a simple and direct means to determine carrier transport properties without the need to evaluate various physical quantities or scan the pump-probe offset. Rather, using tightly focussed Gaussian laser beams for both the pump and probe beams, the modulated reflectance signal is measured as a function of the longitudinal (\emph{Z}) displacement of the sample from focus.  The laser probe allows the modulated component of the reflected beam to be treated by the method of Gaussian decomposition, thus enabling a direct evaluation of transport properties \cite{arXiv}. The simplest focal geometry involves collinear cylindrically symmetric beams directed at normal incidence to the sample.  A microscope objective mounted on a translation stage may then be used to control of the distance of the focal plane from the sample surface.  At each value of \emph{Z}, the modulated component of the retro-reflected probe beam will also be a Gaussian beam with its beam profile determined soley by the focal parameters and the (complex) diffusion length. The reflected probe beam is collected and input to the detector, thereby integrating over the transverse profile of the reflected beam.  This results in an analytic expression for the \emph{Z} dependence of the LPR signal in terms of the focal parameters and the carrier diffusion length.  Best fit values for the diffusion length and recombination lifetime may then be obtained directly from a nonlinear least squares fit procedure \cite{arXiv}. Thus the measurement protocol does not require detailed modeling of electronic transport in the sample (including the numerical solution required in the 3D limit) or an interpretation of the physics underlying the LPR signal.  Furthermore, the treatment by the method of Gaussian decomposition interpolates smoothly between 1D and 3D limits, resulting in the decoupling of diffusion length and recombination lifetime within the fit procedure. The output diffusion lengths and recombination lifetimes and their estimated uncertainties may be combined to provide precision estimates of the diffusion coefficient, or equivalently, the mobility (via the Einstein relation).

\begin{figure}[t!]
\includegraphics[width=246pt]{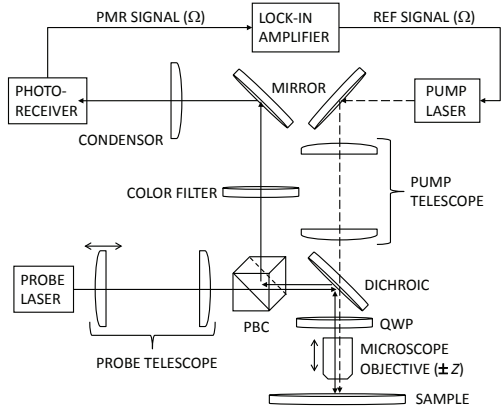}
\caption{\label{fig:system}Schematic of the \emph{Z}-scanning LPR system.}
\end{figure}

In addition to the general PR experimental considerations noted above, the \emph{Z}-scanning LPR technique requires two basic features: (i) a laser probe beam and (ii) the ability to step the sample through focus.    Further, for the reasons discussed below, the spot size at focus should be roughly commensurate with the diffusion length and the modulation period should be roughly commensurate with the recombination lifetime.  The experimental setup is similar to a basic LPR setup previously reported \cite{TSF}. The schematic \emph{Z}-scanning LPR system is shown in Fig.~\ref{fig:system}.  The pump beam is generated by a low power diode laser which is amplitude modulated using, for example, a square wave reference signal from a wide bandwidth lock-in amplifier.  The probe beam is generated by a low power continuous wave diode laser. The pump and probe beams are made collinear through the use of a dichroic beamsplitter and co-focused to a spot on the sample surface using an microscope objective.  The relative focal positions of the pump and probe beams are controlled with independent telescoping lens arrangements in either input beam path.  The intensity of the pump and probe beams at focus may be controlled via neutral density filters fixtured in either input beam path (not shown).  The photo-injected carrier density is preferably maintained in low-injection ($\Delta N/N \ll 1$) such that any variation of recombination lifetime with carrier density may be neglected.  The objective is mounted on a stage which provides for control of the displacement of the common focal plane from the sample. As the pump radiation interacts with the sample, the sample obtains a reflectance modulation, which modulates the reflected probe light intensity.  The incident probe beam is retro-reflected through the microscope objective by the sample.  A polarization beamsplitter operating in conjunction with a quarter wave plate is used to switch the retro-reflected probe beam out of the incoming probe beam path.  The reflected probe beam is projected through a color filter and/or onto a dielectric mirror in order to remove any residual pump light and/or any photo-luminescence signal.  The reflected probe beam is collected at a condenser lens and input to the photoreceiver, thereby integrating over the radial profile of the beam.  The detection path is configured to maintain an ``open aperture'' condition (\emph{i.e.}~such that none of the optical elements clip the reflected probe beam) as the objective is translated with respect to the sample, including the requirement that the active area of the photoreceiver remains underfilled by the reflected probe beam.  In this case the \emph{Z} dependence of the LPR signal will depend solely on the focal parameters and the (complex) carrier diffusion length \cite{arXiv}.  The photoreceiver output signal is passed to the lock-in amplifier, which measures the LPR signal.  The microscope objective is stepped through focus and the open aperture LPR signal is acquired sequentially as a function of \emph{Z}. A nonlinear regression analysis may then be performed to determine the transport properties of the sample according to the theoretical principles described below.

\section{Theoretical Principles}

As noted, near the critical points of a semiconductor material the PR signal exhibits a sharp third-derivative lineshape which arises from an electromodulation effect.  In this case the PR signal becomes:
\begin{eqnarray}
\frac{\Delta R}{R}=\frac{2qN\Delta V}{\epsilon_{s}}\times L(\lambda)
\label{eq:scpr},
\end{eqnarray}
where $q$ is the electronic charge, $\Delta V$ is the photovoltage, $\epsilon_{s}$ is the static dielectric constant, and $L(\lambda)$ is a third-derivative line-shape function determined by the semiconductor bandstructure ($\lambda$ is the probe beam wavelength) \cite{Bottka}.  Eq.~(\ref{eq:scpr}) is valid for depleted interfaces provided the electric field is not too inhomogeneous \cite{Aspnes69,Aspnes73}.  The PR signal is seen to be linearly proportional to the photovoltage. The photovoltage depends on the pump beam and the physics of its interaction with the sample \cite{SPV}.  Based upon junction transport theory \cite{Sze}, the photovoltage takes the simple form:
\begin{eqnarray}
\Delta V = \frac{\eta k_{b}T}{q} \ln \left[\frac{J_{p}}{J_{o} + J_{res}} + 1 \right],
\label{eq:jpv}
\end{eqnarray}
where $\eta$ is an ideality factor, $k_{b}$ is the Boltzmann constant, $J_{p}$ and $J_{o}$ are the photocurrents due to the pump and probe beams, respectively, and $J_{res}$ is the restoring current \cite{Behn}.  The photovoltage is generally linear in the pump intensity provided the photo-injection is small with respect to the restoring current \cite{Schroder97,Park}. In this case the photovoltage will exhibit the spatial dependence of the excess carrier density. In the limit $\omega_{p}(Z)\gg L_{d}$, the photovoltage may be obtained from the solution of the 1D differential equation for the modulated the carrier density.  In the 3D limit ($\omega_{p}(Z)\leq L_{d}$), the excess carrier density involves a Hankel transform of the 1D solution and therefore must be treated numerically. However, the use of laser beams for both the pump and probe means the electric field of the reflected probe beam may instead be treated directly according to the analytically tractable method of Gaussian decomposition \cite{SBahae,Weaire}.

Consider cylindrically symmetric Gaussian pump and probe beams directed at normal incidence onto a sample.  The beams are collinear and co-focused along the \emph{z}-axis.  The displacement of the sample surface from the common beam waist is \emph{Z}. At $Z=0$ the pump will induce a reflectance modulation within a radius of modulation:
\begin{eqnarray}
\omega_{m} \equiv (\omega_{p}^{2}+L_{d}^{2})^{1/2},
\label{eq:rmdef}
\end{eqnarray}
where $\omega_{p}$ is the pump beam waist and $L_{d}$ is the carrier diffusion length. Eq.~(\ref{eq:rmdef}) follows from the 3D differential equation for the photo-injected carrier density generated by a cylindrical Gaussian source \cite{Mandelis}.  Therefore, as the sample is stepped through focus the area of modulation will depend upon the diffusion length according to:
\begin{eqnarray}
\omega_{m}^{2}(Z) \equiv \omega_{p}^{2}(Z)+L_{d}^{2},
\label{eq:rmz}
\end{eqnarray}
where $\omega_{p}(Z) = \omega_{p}\sqrt{1+(Z/z_{p})^{2}}$, and $z_{p}$ is the Rayleigh range of the pump beam (\emph{i.e.} $z_{p}=\pi\omega_{p}^{2}/\lambda_{p}$, where $\lambda_{p}$ is the pump beam wavelength) \cite{Kogelnik}.  The Fresnel coefficient for the reflected probe beam includes the changes due to pump-induced energy transformation processes.  In particular, the electric field of the mirror-reflected probe beam contains a nonlinear phase shift which simply follows the radial modulation profile. The
mirror-reflected probe field may then be decomposed into a sum of Gaussian beams via a Taylor series expansion of the nonlinear phase term \cite{SBahae,Weaire}.  Given a dominant photovoltage effect according to Eq.~(\ref{eq:scpr}), and retaining only two terms in the expansion, the electric field of the reflected probe laser beam at the surface of the sample (disregarding the common spatial phase) may be written:
\begin{eqnarray}
E_{r} = \frac{E_{o}\omega_{o}}{\omega(Z)}\exp \left\{
\frac{-\rho^{2}}{\omega^{2}(Z)} \right\}
\left[\tilde{r} + \frac{\partial\tilde{r}}{\partial n}(n_{2}+ik_{2})
\frac{I_{p} \omega_{m}^{2}}{\omega_{m}^{2}(Z)} \exp \left\{
\frac{-2\rho^{2}}{\omega_{m}^{2}(Z)}\right\} \right]
\label{eq:rpfield},
\end{eqnarray}
where $|E_{o}|^{2}$ is the intensity of the probe beam at focus, $\omega_{o}$ is the probe beam waist (\emph{i.e.} $\omega(Z) = \omega_{o}\sqrt{1+(Z/z_{o})^{2}}$, where $z_{o} = \pi\omega_{o}^{2}/\lambda$ is the Rayleigh range of the probe beam), $\rho$ is radial distance as measured from the probe beam axis, $\tilde{r}$ is the complex reflectance coefficient, $n$ is the refractive index, $I_{p}$ is the intensity of the pump beam at focus, $\omega_{m}(Z)$ is the radius of modulation as defined in Eq.~(\ref{eq:rmz}), and $n_{2}$ and $k_{2}$ are effective nonlinear indices defined by the coefficients appearing in Eq.~(\ref{eq:scpr}) \cite{arXiv}. The leading term corresponds to the DC component of the reflected beam whereas the second term corresponds to its modulated component. The retention of only two terms in the Taylor expansion operates to neglect terms of second order or higher in the nonlinear indices. Phase modulation of the reflected field due to any thermal expansion of the surface has also been neglected since this will not contribute to the open aperture signal \cite{Petrov,Opsal87}.  The modulated component of the reflected probe beam is also a Gaussian beam with its radius defined by the incident probe beam radius and the (diffusion dependent) radius of modulation.  The effect of carrier diffusion on the reflected probe beam field is accounted for by the $\omega_{m}^{2}$ factors appearing in the modulated component.  In particular, the physical fact that the probe beam samples the modulation of the pump beam, as broadened by carrier diffusion, is made explicit by the inclusion of the $\omega_{m}^{2}$ and $\omega_{m}^{2}(Z)$ factors in both the Gaussian exponential and the amplitude prefactor \cite{Kogelnik}.

\begin{figure}[h!]
\includegraphics[width=246pt]{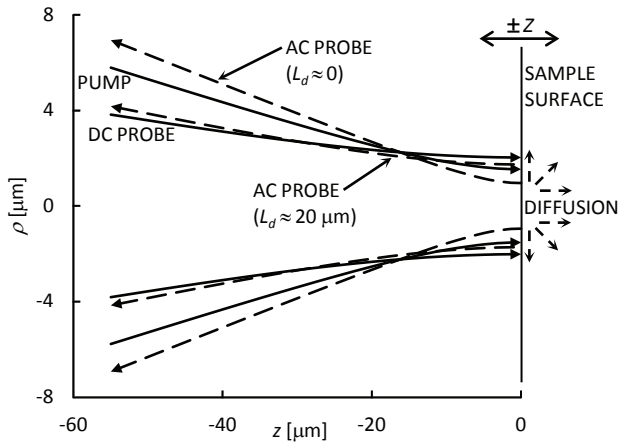}
\caption{\label{fig:focus}Calculated cross-section of beams present in the \emph{Z}-scanning LPR system for $Z=0$.  In the 1D limit the waist of the AC reflected probe beam is smaller than the waist of the DC probe beam.  In the 3D limit the waist of the AC reflected probe beam approaches the waist of the DC probe beam.}
\end{figure}

The focal geometry of the incident pump and probe beams at focus ($Z=0$) is illustrated in Fig.~\ref{fig:focus}. The pump ($\lambda_{p}=488$ nm) and probe ($\lambda=375$ nm) beam waists are $\cong 1.53$ and $\cong 2.02$ $\mu$m, respectively, $z$ is the distance along the common beam axis as measured from the common focal plane, and $\rho$ is the radial distance as measured from the common beam axis. At $Z=0$, the linear reflected pump and probe beam profiles will coincide with their respective input beam profiles. When $L_{d}\approx 0$, the effective waist of the AC reflected probe beam is smaller than the waist of the DC probe beam.  However, in the 3D limit (as illustrated \emph{e.g.}, when $L_{d}\approx 20$ $\mu$m), the AC reflected probe beam approaches the linear reflected probe beam profile.  Thus the profile of the modulated component of the reflected beam is seen to be highly sensitive to carrier diffusion.

Squaring the mirror-reflected probe field and integrating the over the beam profile yields the open aperture LPR signal via the identification:
\begin{eqnarray}
\frac{R + \Delta R}{R}=\frac{\int_{0}^{\infty} |E_{r}|^{2} \rho d\rho}{\int_{0}^{\infty} |E_{dc}|^{2} \rho d\rho},
\label{eq:oaid}
\end{eqnarray}
where $E_{dc}$ is just the linear reflectance amplitude \cite{SBahae}. Neglecting terms of second order in the nonlinear indices and performing the spatial integrations in Eq.~(\ref{eq:oaid}), the open aperture LPR signal becomes:
\begin{eqnarray}
\frac{\Delta R}{R}=\frac{4 n_{2} I_{p}}{n^{2}-1} \times \frac{\omega_{p}^{2}+L_{d}^{2}}{\omega^{2}(Z)+\omega_{p}^{2}(Z)+L_{d}^{2}},
\label{eq:oacalc}
\end{eqnarray}
where $n^{2}\gg k^{2}$.  In particular, the factor $4n_{2}/(n^{2}-1)$ follows from taking 2Re$[(n_{2}+ik_{2})\tilde{r}^{*}d\tilde{r}/dn]/\tilde{r}^{*}\tilde{r}$, where $\tilde{r}=(n+ik-1)/(n+ik+1)$ (and hence $d\tilde{r}/dn=2/(n+ik+1)^{2})$, subject to the condition $n^{2}\gg k^{2}$. Likewise, upon performing the integrals in Eq.~(\ref{eq:oaid}), it is seen:
\begin{eqnarray}
\frac{\Delta R}{R} \propto \frac{\omega_{m}^{2}}{\omega_{m}^{2}(Z)} \times
\frac{\omega_{ac}^{2}(Z)}{\omega^{2}(Z)},
\label{eq:oaints}
\end{eqnarray}
where
\begin{eqnarray}
\frac{1}{\omega_{ac}^{2}(Z)} \equiv \frac{1}{\omega^{2}(Z)} + \frac{1}{\omega_{m}^{2}(Z)}.
\label{eq:wac}
\end{eqnarray}
The ratio $\omega_{m}^{2}/\omega_{m}^{2}(Z)$ comes from the amplitude prefactor associated with the modulated pump beam, whereas the ratio $\omega_{ac}^{2}/\omega^{2}(Z)$ follows from the definite integrals of $xexp\{-x^2\}$ ($=1/2$) proscribed in Eq.~(\ref{eq:oaid}).
However, from Eq.~(\ref{eq:wac}), we have:
\begin{eqnarray}
\frac{\omega^{2}(Z)}{\omega_{ac}^{2}(Z)} & = & 1 + \frac{\omega^{2}(Z)}{\omega_{m}^{2}(Z)} = \frac{\omega_{m}^{2}(Z) + \omega^{2}(Z)}{\omega_{m}^{2}(Z)}.
\label{eq:winv}
\end{eqnarray}
Substituting Eq.~(\ref{eq:winv}) into Eq.~(\ref{eq:oaints}) and cancelling the factors of $\omega_{m}^{2}(Z)$ yields:
\begin{eqnarray}
\frac{\Delta R}{R} \propto \frac{\omega_{m}^{2}}{\omega^{2}(Z) + \omega_{m}^{2}(Z)},
\label{eq:zdep}
\end{eqnarray}
in accord with Eq.~(\ref{eq:oacalc}). It should be noted the dependence of $\Delta R$ on \emph{Z} is indentical to the dependence shown in Eq.~(\ref{eq:zdep}) since in this case the $1/\omega^{2}(Z)$ term of Eq.~(\ref{eq:oaints}) arises from the amplitude prefactor appearing in the linear component of the mirror-reflected probe field.  Thus the \emph{Z} dependence of the LPR signal as shown in Eq.~(\ref{eq:oacalc}) is contained entirely within $\Delta R$. (This is intuitively obvious since the open aperture condition necessarily eliminates any \emph{Z} dependence of $1/R$).  It also may be shown Eq.~(\ref{eq:oacalc}) holds when the effect of photo-injection by the probe beam on the built-in electric field is included in the analysis.

Thus integrating over the radial dependence of the reflected probe beam results in a simple analytic expression for the \emph{Z} dependence of the LPR signal. In particular, it is seen the \emph{Z}-profile of the LPR signal depends soley on the focal parameters and the diffusion length and is contained entirely within the denominator of Eq.~(\ref{eq:oacalc}).  The appearance of $L_{d}^2$ in the denominator of Eq.~(\ref{eq:oacalc}) shows the \emph{Z}-profile of the LPR signal will depend strongly on diffusion length provided the pump and probe beam waists are commensurate with $L_{d}$. Moreover, well away from $Z = 0$ (\emph{i.e.}~where $\omega^{2}(Z)+\omega_{p}^{2}(Z) \gg L_{d}^{2}$), the 1D limit is restored.  Thus the treatment by the method of Gaussian decomposition smoothly interpolates between the 3D and 1D limits.

At intermediate frequencies where the recombination lifetime $\tau$ is comparable to the modulation period, $\tau$ likewise becomes coupled into the \emph{Z} dependence of the LPR signal through the appearance of the complex diffusion length $\tilde{L}_{d} \equiv L_{d}/\sqrt{1+i\Omega\tau}$, where $\Omega$ is the modulation frequency in radians per second. In particular, Eq.~(\ref{eq:oacalc}) demonstrates that the LPR signal as a function of \emph{Z} may be parameterized by the expression:
\begin{eqnarray}
\frac{\Delta R}{R}=\frac{A\exp i\phi_{o}}{\omega^{2}(Z)+\omega_{p}^{2}(Z)+\tilde{L}_{d}^{2}},
\label{eq:polar}
\end{eqnarray}
where $A$ and $\phi_{o}$ are the \emph{Z}-independent amplitude and phase, respectively \cite{arXiv}.  As noted the PR phase characterizes the time lag of the modulated reflectivity response due to the non-instantaneous interaction dynamics of carrier within the sample (and therefore should not be confused with the optical phase). While the LPR signal generally depends upon carrier diffusion through the dependence of $\Delta V$ (or $\Delta N$) on $L_{d}$ and/or $\tau$, these dependencies are absorbed into $A$ and $\phi_{o}$ and therefore do not directly enter into the \emph{Z} dependence of the signal.  Furthermore, in the case of stratified media and provided interference effects can be neglected, the \emph{Z}-scanning LPR signal will take the form of a series of terms of the form of Eq.~(\ref{eq:polar}) wherein the diffusion length appearing in each term is the just the diffusion length in the associated modulating layer \cite{Aspnes73,TSF}. Thus the \emph{Z} dependence of the LPR signal does not directly depend upon carrier transport in the direction normal to the sample surface.

The coupling of $\tilde{L_{d}}$ into the \emph{Z} dependence of the LPR signal indicates that $L_{d}$, and ultimately $\tau$, may be determined by a regressive fit to the experimental \emph{Z}-scanning LPR data.
For example, the polar decomposition of Eq.~(\ref{eq:polar}) yields two nonlinear equations containing a number of common parameters including the diffusion length and recombination time.  Complex nonlinear least squares optimization is the preferred method to fit small-signal AC response data such as light-scattering complex data \cite{MacDonald92}. In this approach, a model representative of the response is simultaneously fit to the real and imaginary components of the data, yielding single set of parameter estimates based on all the data \cite{BEchem,Boukamp,MacDonald87}. However, to optimize the overall fit, the proper choice of weighting between real and imaginary fits is crucially important \cite{MacDonald92}. Rather than go into such details here, it is more illuminating for present purposes to analyze the amplitude and phase of Eq.~(\ref{eq:polar}) separately.

The analytic expression for the \emph{Z}-dependence of the LPR amplitude is given by:
\begin{widetext}
\begin{eqnarray}
\left| \frac{\Delta R}{R} \right| = \frac{A}{\sqrt{L_{d}^{4} + 2L_{d}^{2}[\omega_{p}^{2}(Z) + \omega^{2}(Z)] + [\omega_{p}^{2}(Z) + \omega^{2}(Z)]^{2}(1 + \Omega^{2}\tau^{2})}}.
\label{eq:amp}
\end{eqnarray}
\end{widetext}
This expression may be parameterized by the set of variables: $A$, $L_{d}^{2}$, $\omega_{p}^{2}$, $\omega_{o}^{2}$, and $\Omega\tau$.  In principle, a standard nonlinear regression analysis can then be used to adjust the variables within the parameterized expression for the amplitude to provide an optimum fit \cite{NumRec}. For analysis purposes, the system parameters ($\Omega$, $\omega_{o}$, and $\omega_{p}$) may be treated as fixed, whereas the sample parameters ($A$, $\phi_{o}$, $L_{d}^{2}$, and $\tau$) may be treated as variable parameters to be resolved by the fit procedure.  The parameters $A$, $L_{d}^{2}$, and $\Omega\tau$ are generally correlated within the amplitude fit. However, by inspection the $\Omega\tau$ dependence of the amplitude only enters in the 1D tails of the data. In this limit, the amplitude becomes:
\begin{eqnarray}
\left| \frac{\Delta R}{R} \right| \simeq \frac{A}{[\omega^{2}(Z) + \omega_{p}^{2}(Z)]\sqrt{1 + \Omega^{2}\tau^{2}}}.
\end{eqnarray}
Conversely, in the 3D limit the amplitude becomes:
\begin{eqnarray}
\left| \frac{\Delta R}{R} \right| \simeq \frac{A}{L_{d}^{2}} \times \left[ 1 - \frac{\omega_{p}^{2}(Z) + \omega^{2}(Z)}{L_{d}^{2}} + \cdots \right].
\end{eqnarray}
Thus, at intermediate frequencies and provided the strong 3D limit is realized ($L_{d}^{2} \gg \omega_{p}^{2}(Z) + \omega^{2}(Z)$), the amplitude will become independent of $\Omega\tau$.  This decoupling of $L_{d}^{2}$ and $\Omega\tau$ in the respective 3D and 1D limits suggests the amplitude alone may be used  to determine both the diffusion length and the recombination time.  However in practice, the mild functional dependence upon $\Omega\tau$ (in the tails of the data) limits the ability to determine recombination time from the amplitude alone.

On the other hand, the expression for the phase provides a reliable (and surprising) means to extract the recombination time. The analytic expression for the \emph{Z} dependence of the LPR phase is given by:
\begin{eqnarray}
\phi = \phi_{o} + \arctan \left\{\frac{L_{d}^{2}\Omega\tau}{{L}_{d}^{2} + [\omega_{p}^{2}(Z) + \omega^{2}(Z)](1 + \Omega^{2}\tau^{2})}\right\}.
\label{eq:phase}
\end{eqnarray}
In general, the phase in a modulated photovoltage measurement will exhibit an arctangent dependence approaching zero for $\Omega\tau \ll 1$, and $-90^{\circ}$ for $\Omega\tau \gg 1$ \cite{Park}.  Thus $\phi_{o}$ always takes a negative value between $0$ and $-\pi/2$. The second term on the r.h.s.~of Eq.~(\ref{eq:phase}) represents the 3D correction to the 1D phase and is always positive.  Thus the 3D correction always reduces the absolute value of the phase.

The phase expression allows parametrization using the variables: $\phi_{o}$, $L_{d}^{2}$, $\omega_{p}^{2}$, $\omega_{o}^{2}$, and $\Omega\tau$.  While the parameters $\phi_{o}$, $L_{d}^{2}$, and $\Omega\tau$ are generally correlated in the phase fit, in the 3D limit it may be shown the phase becomes:
\begin{eqnarray}
\phi \simeq \phi_{o} + \arctan \{\Omega\tau\} - \left[\frac{\omega_{p}^{2}(Z) + \omega^{2}(Z)}{L_{d}^{2}}\right] \Omega\tau + \cdots,
\end{eqnarray}
whereas in the 1D limit the phase is:
\begin{eqnarray}
\phi \simeq \phi_{o} + \left[\frac{L_{d}^{2}}{\omega_{p}^{2}(Z) + \omega^{2}(Z)}\right] \frac{\Omega\tau}{1 + \Omega^{2}\tau^{2}} - \cdots.
\end{eqnarray}
In either limit, the phase is seen to depend on the ratio $[\omega_{p}^{2}(Z) + \omega^{2}(Z)]/L_{d}^{2}$.  However, in the 3D limit the phase will approach $\phi_{o} + \arctan\{\Omega\tau\}$, whereas in the 1D limit the phase will approach $\phi_{o}$ (\emph{i.e.} $\phi_{o}$ is just the ``1D phase'').    The difference in the respective phase limits implies:
\begin{eqnarray}
\Omega\tau = \tan \{\Delta\phi\},
\end{eqnarray}
where $\Delta\phi$ is just the total variation observed in the phase data.   Thus provided the strong 3D and 1D limits are realized the recombination time may essentially be read off the raw phase data.  Note the dependencies of $\phi_{o}$ on various parameters such as $L_{d}$, $\tau$, and/or surface or interface recombination velocities are completely irrelevant to this conclusion. This vastly simplifies the problem of determining the recombination time and illustrates the power of the \emph{Z}-scanning LPR technique.

While the 1D limit may be realized without difficulty, the decreasing signal in the tails of the data will increase the relative experimental uncertainty in this limit.  (Note that due to the open aperture configuration, the measurement noise remains roughly constant as the microscope objective is stepped through focus.  This mitigates the competition between sensitivity and signal to noise encountered in offset pump-probe techniques \cite{Idaho}.) On the other hand, the strong 3D limit may not be attainable due to the spatial resolution (as determined by the value of $\omega_{p}^{2} + \omega_{o}^{2}$).  Nevertheless, it should be clear from the foregoing that the diffusion length primarily impacts the amplitude data while the recombination time primarily impacts the phase data.  These observations suggest an iterative procedure involving independent nonlinear fits to the amplitude and phase expressions in order to establish $L_{d}$ and $\tau$ and their statistical uncertainties.  For example, the amplitude data may be first fit with $\Omega\tau$ fixed ($\sim 1$) to determine a reasonably accurate value for $L_{d}^{2}$ and its uncertainty \cite{NumRec}. Then the output value for $L_{d}^{2}$ may be held constant in the phase fit in order to estimate $\Omega\tau$ and its uncertainty. This procedure can be iterated, holding each $L_{d}^{2}$ value output from the amplitude fit constant in the subsequent phase fit, and each $\Omega\tau$ value output from the phase fit constant in the subsequent amplitude fit, until $L_{d}^{2}$ and $\Omega\tau$ approach limiting values.  Also, since the measurement noise remains roughly constant as the microscope objective is stepped through focus, the estimated uncertainties of the output fit parameters depend primarily upon the number and spacing of the data points in \emph{Z} \cite{NumRec}. Thus the polar decomposition of the \emph{Z}-scanning LPR signal provides a straightforward means to evaluate $L_{d}$ and $\tau$ with high precision. The mobility $\mu$ and its uncertainty may then be obtained from the Einstein relation $\mu = qD/k_{b}T$, where $D \equiv L_{d}^{2}/\tau$ is the ambipolar diffusion coefficient and $k_{b}T/q$ is the thermal voltage ($\cong26$ mV)

\section{Results and Discussion}

In order to illustrate the practical use of the technique to determine carrier transport properties, a set of silicon samples with \emph{p-n} junctions formed by advanced ion implantation and annealing processes were evaluated.  The use of LPR to characterize electrical activation in these samples (\emph{i.e.}~via the dependence of Eq.~(\ref{eq:scpr}) on $N$) has previously been reported \cite{JVST}.  The initial disclosure of the \emph{Z}-scanning LPR technique also utilized a subset of these samples \cite{arXiv}.  To form the \emph{p-n} junctions, $10-15$ $\Omega$-cm \emph{n}-type silicon (100) substrates were implanted with As (\emph{n}-type dopant) followed by low-energy high-dose B (\emph{p}-type dopant) implantation.  Dopant activation was performed using millisecond timescale flash-lamp based annealing. A range of base temperature and flash temperature targets were used to study the effect of process conditions on dopant activation, implanted dopant diffusion, and material quality.  The process conditions evaluated included: (i) flash target temperatures in the $1250-1350^{\circ}$C range, (ii) an additional thermal annealing of the As counter-doping layer prior to B implantation, and (iii) use of a Ge amorphizing implant (AI).  The primary purpose of the AI process was to reduce ion channeling during the subsequent B implant via the introduction of crystalline defects close to the sample surface.  These defects also reduce the carrier diffusion length and recombination time in the implanted region. SIMS data indicated post-activation B doping levels of $\approx 1\times 10^{19}$/cc at $X_{j}\approx 20$ nm across the sample set.  The conventional 1D Poisson analysis \cite{Sze} estimates the junction voltages on the order of several volts.

In addition to the system details as described above, the LPR setup utilized here incorporated an avalanche photodiode (APD) detector (Hamamatsu model C12703 high-gain APD module configured with the Hamamatsu model S5344 short wavelength APD) which allowed the reduction of the probe beam power by three (3) orders of magnitude.  This reduces the photocurrent generated by the probe and therefore extends the range of photovoltages detectable by the system in accord with Eq.~(\ref{eq:jpv}).  However, this feature is relatively unimportant for the \emph{p-n} junctions evaluated here since the photo-injected carrier densities remain small with respect to the restoring current \cite{JVST,TSF}. The pump and probe wavelengths were $488$ and $374$ nm, respectively. The probe beam wavelength is near the $E_{1}$ critical point in Si, resulting in a dominant photovoltage effect. The absorption depth of the probe is $\delta \cong 23$ nm.  Thus any detected photovoltage must occur at or near the surface. The pump output was amplitude modulated via a reference signal from a wide bandwidth lock-in amplifier (Stanford Research Systems model SR844).  As also noted, the phase in a modulated photovoltage measurement will exhibit an arctangent dependence approaching zero for $\Omega\tau \ll 1$, and $-90^{\circ}$ for $\Omega\tau \gg 1$. Thus, in the intermediate regime, the phase will exhibit linearity (centered about the value $\phi = -45^{\circ}$).  For each of the wafers used in this study, this linearity was confirmed over the frequency range $\sim 500-850$ kHz.  Thus an operational modulation period of $660$ kHz was selected.  \emph{Z}-scanning LPR data was acquired from the samples.  Provisional estimates for $L_{d}^{2}$ were obtained via regressive fitting to the amplitude data. Then the values for $L_{d}^{2}$ were held constant in a fit to the phase data in order to estimate $\Omega\tau$. This procedure was iterated until $L_{d}^{2}$ and $\Omega\tau$ approached limiting values. The estimated uncertainties in the extracted parameters were also output from the fitting procedure.

\begin{figure}[t!]
\includegraphics[width=246pt]{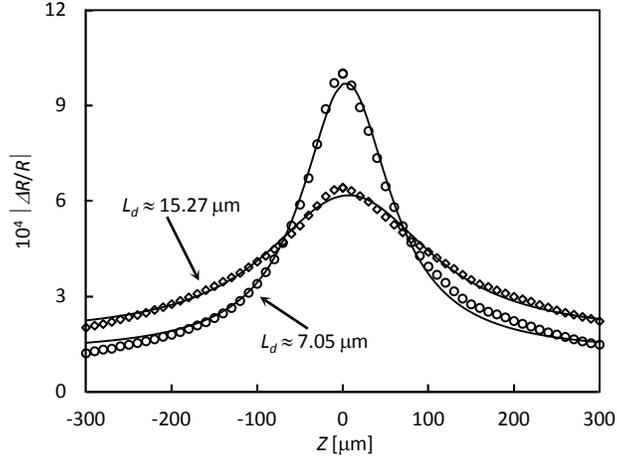}
\caption{\label{fig:Lfit}\emph{Z}-scanning LPR amplitude data and fits showing the effect of near surface damage (due to AI) on shallow electrical junctions formed in silicon.  The more narrow \emph{Z}-profile indicates a shorter diffusion length.}
\end{figure}

\begin{figure}[t!]
\includegraphics[width=246pt]{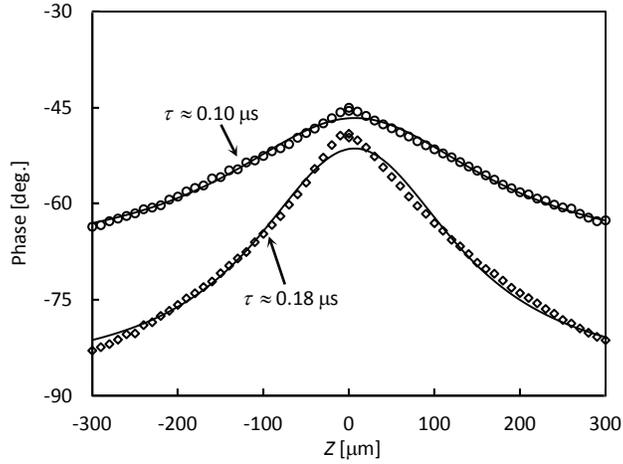}
\caption{\label{fig:taufit}\emph{Z}-scanning LPR phase data from the same pair of samples as shown in Fig.~\ref{fig:Lfit}, again showing the effect of near surface damage (due to AI) on the junction.  The broader \emph{Z}-profile indicates a shorter recombination lifetime.}
\end{figure}

Fig.~\ref{fig:Lfit} shows experimental \emph{Z}-scanning LPR amplitude data and fits obtained from samples with and without AI. The \emph{Z}-scan data increment was $10$ $\mu$m, with a maximum \emph{Z} displacement of $\pm 300$ $\mu$m. The amplitudes are symmetric with respect to \emph{Z}, as anticipated. The amplitude from the sample without AI shows a relatively broad \emph{Z}-profile, whereas the data from the sample with AI exhibits a narrower profile.  The more sharply peaked LPR response as a function of \emph{Z} seen on the sample with AI indicates a shorter diffusion length.  This behavior was apparent in the raw data for all samples that received the AI process. Thus the effect of AI on carrier diffusion is strikingly evident in the raw amplitude data.  Fig.~\ref{fig:taufit} shows experimental phase data and fit obtained from the same pair of samples as shown in Fig.~\ref{fig:Lfit}. The phases are again symmetric with respect to \emph{Z}, in accord with Eq.~(\ref{eq:phase}).  The observed phase lag $\sim -45^{\circ}$ confirms operation in the intermediate frequency regime. The broader phase as a function of \emph{Z} seen on the sample with AI evidences a shorter recombination lifetime.  The more sharply peaked amplitude data corresponds to the broader phase data, indicating carrier relaxation in the sample with AI happens more quickly and occurs over a shorter range than in the sample without AI.  This behavior was apparent in the raw phase data for all samples that received the AI process.  As the \emph{p-n} junction samples considered here were nominally identical beyond a depth $\sim 60$ nm, the remarkable impact of near surface damage on the \emph{Z}-profiles (as seen in Figs.~\ref{fig:Lfit} and \ref{fig:taufit}) also confirms the $374$ nm probe is primarily sensitive to near-surface carrier transport.  This near-surface specificity is a key advantage in advanced semiconductor manufacturing. The mobility and its estimated uncertainty were then obtained from the extracted parameters via the Einstein relation.

\begin{table*}[h!]
\centering
\caption{\label{tab:table1} Measured carrier diffusion lengths, recombination times and mobilities, as determined via fitting to \emph{Z}-scanning LPR data obtained from the subset of samples formed using AI.}
\begin{tabular}{ccccccc}
\hline
&\multicolumn{2}{c}{$L_{d}$ [$\mu$m]}&\multicolumn{2}{c}{$\tau$ [ns]}&\multicolumn{2}{c}{$\mu$ [$\mathrm{cm}^{2}/\mathrm{V}\cdot\mathrm{s}$]}\\
Flash temp [$^{\circ}$C]&Flash only&As pre-soak&Flash only&As pre-soak&Flash only&As pre-soak\\
\hline
$1250/550$&
$7.84\pm 0.03$& $9.09\pm 0.03$&
$151.9\pm 0.5$& $165.1\pm 0.6$&
$155\pm 2$& $192\pm 2$\\

$1300/550$&
$5.47\pm 0.01$& $5.36\pm 0.01$&
$122.5\pm 0.7$& $131.8\pm 0.7$&
$94\pm 1$& $84\pm 1$\\

$1300/550(2X)$&
$6.54\pm 0.01$& $7.05\pm 0.01$&
$99.9\pm 0.3$& $99.5\pm 0.6$&
$165\pm 1$ & $192\pm 1$\\

$1300/600$&
$8.41\pm 0.02$& $9.81\pm 0.02$&
$165.2\pm 0.5$& $168.5\pm 0.5$&
$165\pm 1$& $220\pm 1$\\

$1350/600$&
$8.87\pm 0.01$& $10.99\pm 0.01$&
$163.4\pm 0.5$& $168.9\pm 0.5$&
$185\pm 1$& $275\pm 1$\\
\hline
\end{tabular}
\end{table*}

Table~\ref{tab:table1} lists fitted values of diffusion length, recombination time, and mobility for the \emph{p-n} junction samples formed with AI, assuming a measurement uncertainty of $2$ ppm for the LPR amplitude and $0.13^{\circ}$ for the LPR phase. Systematic variations in the measured carrier parameters with process conditions are observed.  Overall, the measured parameters show little sensitivity to the As thermal anneal (columns labeled ``As pre-soak'').  This is expected since the AI step occurred after the As thermal anneal (prior to B implantation). When the target temperature of the flash anneal is increased from $1250^{\circ}$C to $1300^{\circ}$C, the diffusion length, recombination lifetime, and mobility all decrease. This indicates a migration of the AI damage toward the surface (\emph{i.e.}~that only partial recrystallization is achieved). When the $1300^{\circ}$C$/550^{\circ}$C flash anneal is repeated, the diffusion length increases by a factor of $\approx 1.2X$, indicating the onset of junction activation.  However, the recombination time decreases further, indicating further concentration of the AI damage toward the surface. When the base temperature of the flash anneal is increased to $600^{\circ}$C, the recombination time jumps by a factor of $\approx 1.7X$ (with respect to the repeated $1300^{\circ}$C$/550^{\circ}$C anneal), confirming this higher base temperature results in better removal of the AI damage. The observed diffusion length also increases by $\approx 30\%$ indicating better junction activation.  Finally, when the target temperature of the flash anneal is increased to $1350^{\circ}$C, the diffusion length increases by another $\approx 10\%$, but without further increase in recombination time.  This limiting behavior indicates the $1350^{\circ}$C$/600^{\circ}$C flash anneal achieves good junction activation and substantially removes the AI damage.

\begin{table*}[h!]
\centering
\caption{\label{tab:table2} Measured carrier diffusion lengths, recombination times and mobilities, as determined via fitting to \emph{Z}-scanning LPR data obtained from the subset of samples formed without AI.}
\begin{tabular}{ccccccc}
\hline
&\multicolumn{2}{c}{$L_{d}$ [$\mu$m]}&\multicolumn{2}{c}{$\tau$ [ns]}&\multicolumn{2}{c}{$\mu$ [$\mathrm{cm}^{2}/\mathrm{V}\cdot\mathrm{s}$]}\\
Flash temp [$^{\circ}$C]&Flash only&As pre-soak&Flash only&As pre-soak&Flash only&As pre-soak\\
\hline
$1250/550$&
$15.17\pm 0.04$& $21.97\pm 0.05$&
$197.7\pm 0.6$& $269.4\pm 0.6$&
$447\pm 3$& $689\pm 5$\\

$1300/550$&
$15.44\pm 0.02$& $ - $&
$203.6\pm 0.5$& $ - $&
$450\pm 3$& $ - $\\

$1300/550(2X)$&
$20.04\pm 0.01$& $19.09\pm 0.02$&
$285.2\pm 0.9$& $207.0\pm 0.7$&
$542\pm 2$ & $677\pm 3$\\

$1350/600$&
$14.09\pm 0.01$& $15.27\pm 0.02$&
$180.8\pm 0.5$& $179.8\pm 0.5$&
$422\pm 2$& $499\pm 2$\\
\hline
\end{tabular}
\end{table*}

Table~\ref{tab:table2} lists fitted values of diffusion length, recombination time, and mobility for the set of \emph{p-n} junction samples formed without AI (again assuming a measurement uncertainty of $2$ ppm for the LPR amplitude and $0.13^{\circ}$ for the LPR phase). Systematic variations in the measured carrier parameters with process conditions are again observed.  In this case, the As thermal anneal results in notable differences in carrier transport properties, particularly for the lower thermal budget flash anneals.  The anticipated result of the As thermal anneal was full activation of the \emph{n}-type counter-dopant prior to B implantation accompanied by complete recrystallization of any damage from the heavy As implant. Here, since these samples had no subsequent AI to reintroduce damage, recrystallization by the flash anneal is not pertinent. For the target temperature of $1250^{\circ}$C, the diffusion length, recombination lifetime, and mobility are each observed to be $\approx 40\%$ greater for the junction formed with the As thermal anneal. For the flash only process no significant differences in transport properties are observed up to a target temperature of $1300^{\circ}$C. Note the determined mobility for the flash only process is agrees closely with reported values for hole mobility in undoped silicon.  This corresponds to the minority hole mobility in the modulated \emph{n}-type depletion region just below the (one-sided) junction. (Mobility values in excess of the undoped hole mobility in Si, as seen \emph{e.g.}~in the samples with the soak anneal, generally involve a contribution from the electron mobility to the ambipolar mobility.) When the $1300^{\circ}$C$/550^{\circ}$C flash anneal is repeated, the diffusion length and recombination time for the flash only process jumps by $\approx 35\%$, again indicating the onset of junction (B) activation at this thermal budget.  And finally, when the flash temperature is increased to $1350^{\circ}$C$/600^{\circ}$, the carrier transport properties converge.  This limiting behavior indicates the thermal budget at which the flash only process achieves junction activation equivalent to the process incorporating the soak anneal.

It should be emphasized the foregoing interpretations of the effect of process conditions on carrier transport properties are not required to evaluate any transport property, but instead follow from knowledge of the process conditions in view of the determined transport properties.  In particular, the transport properties are directly evaluated using the \emph{Z}-scanning LPR technique. Further, while no comparison with transport values measured via other methods has been made in this work, the principles of the technique demonstrate it provides a self-contained and self-consistent method to establish carrier diffusion lengths, recombination lifetimes, and mobilities. Thus it is unsurprising the values of carrier transport properties reported here are in good agreement with literature values corresponding to the active doping levels (or otherwise expected in view of the process conditions).  Moreover, the ability to perform a regressive fit to the data enables the technique to achieve high precision. For example, the estimated uncertainties in the determined mobility values remain less than $1.2\%$ in all cases.  Thus the technique provides a practical tool for the non-destructive characterization of electronic transport properties.

\section{Conclusions}

The \emph{Z}-scanning LPR technique has been used to quickly and precisely measure carrier transport properties using data obtained from a convenient optical setup.  The technique is based upon profiling of the output signals of a LPR system as the sample is stepped through focus. The only experimental modifications to the conventional PR configuration are the incorporation of a probe laser beam and the ability to step the sample through focus. As the sample is stepped through focus the intensity modulated pump beam will induce a reflectance modulation within an area of defined by the pump spot size and the (complex) diffusion length.  The use of a probe laser beam permits an analytic parameterization for the \emph{Z} dependence of the LPR signal which depends soley on the focal parameters and the diffusion length.  This enables the use of a nonlinear least squares fit procedure to directly analyze transport properties without detailed modeling of transport in the sample.   The precision of the technique depends primarily upon the number and spacing of the data points in \emph{Z}. Further work involving application of complex nonlinear least squares fit techniques to \emph{Z}-scanning LPR data is in progress.

\end{document}